\documentclass[]{spie}

\usepackage[mathlines]{lineno}

\usepackage{graphicx}
\usepackage{tabularx}
\usepackage[caption=false]{subfig}
\usepackage{comment}
\usepackage{siunitx}
\usepackage{mathtools}
\usepackage{color}
\usepackage[percent]{overpic}
\usepackage{tikz}
\usetikzlibrary{backgrounds}
\usepackage{pict2e}

\newcounter{pcounter}
\newcolumntype{C}[1]{>{\centering\let\newline\\\arraybackslash\hspace{0pt}}m{#1}}
\newcolumntype{L}[1]{>{\raggedleft\let\newline\\\arraybackslash\hspace{0pt}}m{#1}}
\newcolumntype{R}[1]{>{\raggedright\let\newline\\\arraybackslash\hspace{0pt}}m{#1}}

\begin{document}


\title{Reconstruction Algorithm Design for Mitigating the Orientation Dependent Conspicuity of Fiber-Like signals in Digital Breast Tomosynthesis} 
\author{Sean D. Rose, Ingrid Reiser, Emil Y. Sidky, and Xiaochuan Pan
\skiplinehalf
The University of Chicago Dept. of Radiology MC-2026, 5841 S. Maryland Avenue, Chicago IL, 60637}

\maketitle 

\begin{abstract}
There are a number of clinically relevant tasks in digital breast tomosynthesis (DBT) involving the detection and visual assessment of fiber-like structures, such as Cooper's ligaments, blood vessels, and spiculated lesions. Such structures can exhibit orientation dependent variations in conspicuity\cite{MaidmentRSNA2017}. This study demonstrates the presence of in-plane orientation-dependent signal conspicuity for fiber-like signals in DBT and shows how reconstruction algorithm design can mitigate this phenomenon. We uncover a tradeoff between minimizing orientation-dependence and preserving depth resolution that is dictated by the regularization strength employed in reconstruction. 
\end{abstract}

\begin{figure*}[b!]
	\centering
	\includegraphics[width=1.0\textwidth]{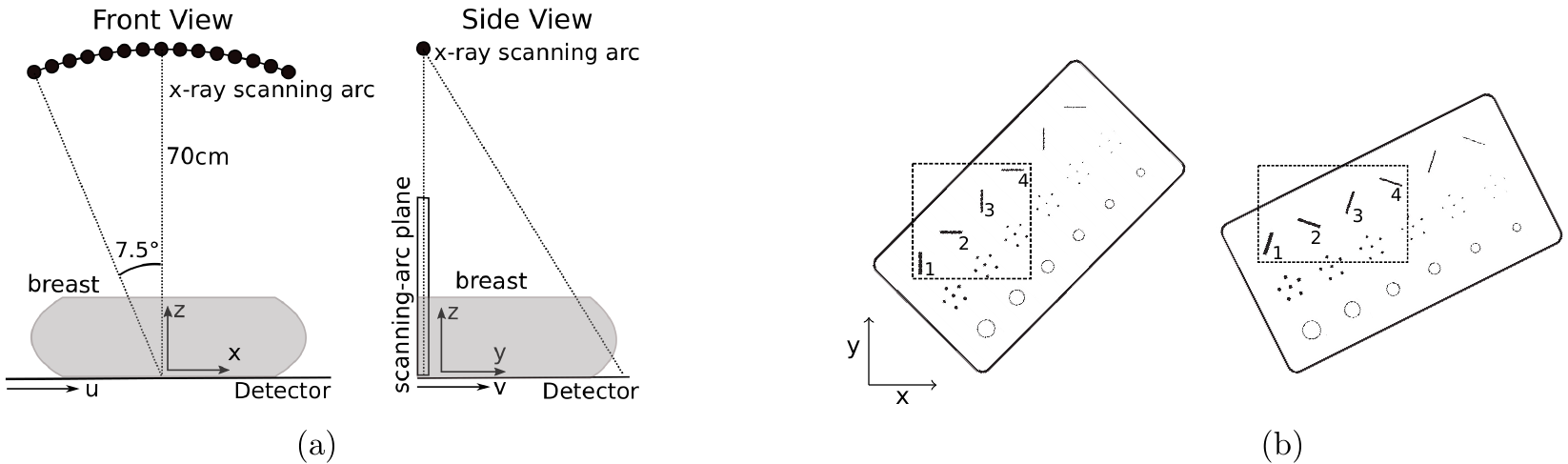}
	\caption{\label{fig1} a) Schematic illustrating DBT scanner configuration. b) Two orientations of the ACR phantom used for assessing orientation dependence. ROIs to be displayed are indicated by dashed boxes. The $x$ direction indicates the direction of source travel.}
\end{figure*}

\section{Purpose}
In DBT, the conspicuity of thin, fiber-like signals exhibits in-plane orientation dependence. The purpose of this work is to investigate the potential of reconstruction algorithm design to mitigate this effect.

\begin{figure*}[t!]
	\centering
	\includegraphics[width=0.99\textwidth]{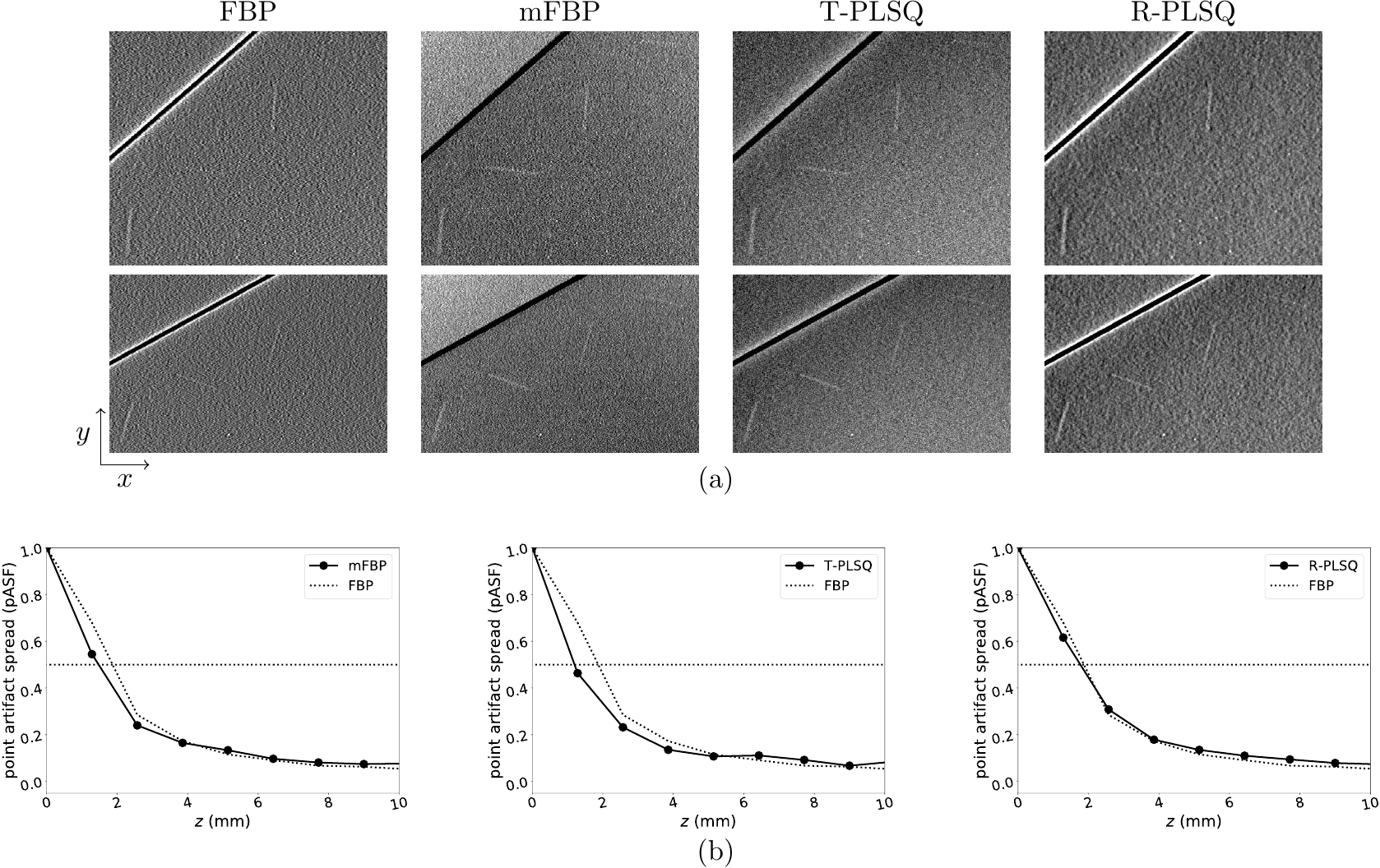}
	\caption{\label{fig2} a) Reconstructions of the ACR phantom using FBP, mFBP, T-PLSQ, and R-PLSQ with the same regularization strengths
employed to calculate the pASFs shown in panel d). Fibers 2 and 4, being oriented more closely to the $x$-axis, exhibit reduced conspicuity in the FBP and R-PLSQ reconstructions.  b) Artifact spread functions of a point-like signal (pASFs) for FBP, mFBP, T-PLSQ, and R-PLSQ. A dashed horizontal line is drawn at a value of 0.5 for reference.}
\end{figure*}

\section{Methods}
Four image reconstruction algorithms are investigated: filtered backprojection (FBP) with Hanning window apodization, a modified implementation of FBP employing a boost to the low-frequency response of the ramp filter (mFBP),\cite{Orman2006,Sechopoulos2013} Tikhonov penalized least squares (T-PLSQ), and roughness penalized least squares (R-PLSQ). \cite{Rose2017} Data was acquired from an ACR digital mammography accreditation phantom at two orientations. A schematic of the scanner's geometry and the two orientations of the phantom are shown in Figure \ref{fig1}. Reconstructions of the phantom were used to compare fiber-like signal conspicuity between algorithms. 

To generate matched resolution reconstructions, a simulation study was performed in which a point-like signal was forward projected and reconstructed using each of the four algorithms. The artifact spread function \cite{Hu2008} of the point-like signal was used to quantify depth resolution. As the artifact spread function is often applied to extended objects, we denote this metric by pASF to emphasize it is specific to a point-like signal.

\begin{figure*}[t!]
	\centering
	\includegraphics[width=0.92\textwidth]{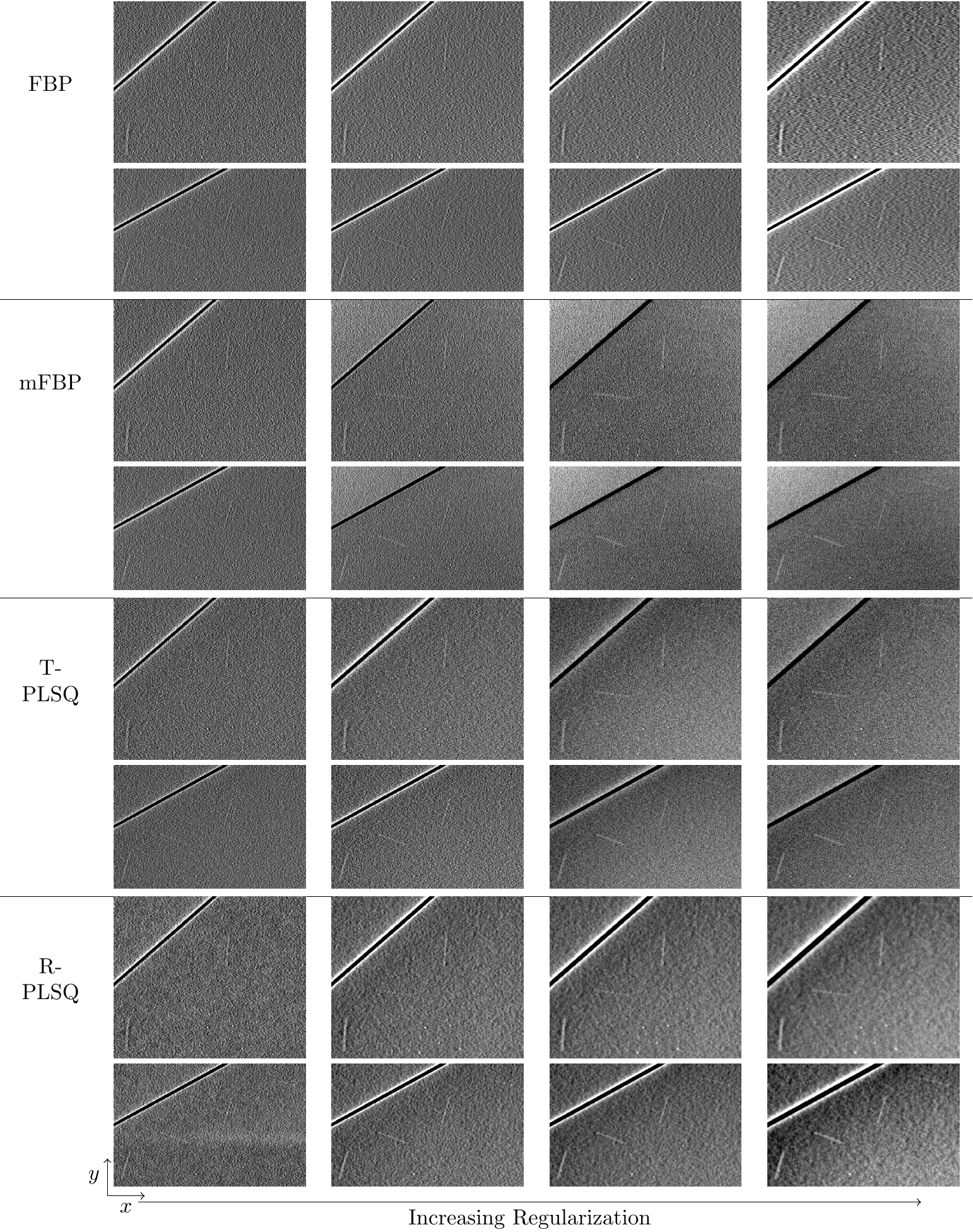}
	\caption{\label{fig3} Reconstructions of the ACR phantom over a range of regularization strengths, increasing from left to right, with each of the four algorithms. Fibers 2 and 4, being more closely aligned with the $x$-axis, exhibit reduced conspicuity at low regularization strengths in each of the four algorithms. As regularization strength is increased, conspicuity of the fibers becomes more in line with fiber diameter.}
\end{figure*}
\section{Results}

In panel a) of Figure \ref{fig2} we show matched depth resolution reconstructions of the four largest diameter fibers of the ACR phantom. The same regularization strengths were used to calculate the pASFs shown in panel d). The horizontal ($x$) axis is the direction of source travel. Note that fibers 2 and 4, when oriented parallel to the $x$ axis, exhibit reduced conspicuity in the FBP reconstruction. Their conspicuity is slightly improved in the R-PLSQ reconstruction and further improved with mFBP and T-PLSQ. This indicates mFBP, T-PLSQ, and R-PLSQ have less orientation dependence than FBP at matched depth resolution. 

The artifact spread function of a point-like signal (pASF) was calculated in simulation and used to quantify the depth resolution of the four reconstruction algorithms. The result is shown in panel b). The reference FBP reconstruction employed a Hanning apodizing window with a cutoff frequency of 0.7$f_{nyq}$, where $f_{nyq}$ is the Nyquist frequency of the detector. This value was chosen based on subjective visualization. The regularization strengths of the other three algorithms were chosen to match the pASF of the FBP reconstruction.

In Figure \ref{fig3} we show reconstructions of the ACR phantom at different regularization strengths for each of the four algorithms. Conspicuity of the second and fourth fibers, when oriented parallel to the $x$-axis, is reduced at low regularization strengths (left) in both the analytic (FBP and mFBP) and iterative (T-PLSQ and R-PLSQ) algorithms. As regularization strength is increased (right), conspicuity of the fibers increasingly tracks with fiber thickness. As increasing regularization strength also deteriorates depth resolution of the reconstruction, this result demonstrates a tradeoff between minimizing orientation-dependence of fiber-like signal conspicuity and preserving depth resolution.

\section{Conclusions}
The regularization strength employed in DBT image reconstruction not only controls noise level and depth resolution, but also the orientation dependence of fiber-like signal conspicuity. All three alternative algorithms designs were seen to exhibit a better tradeoff between orientation dependence and depth resolution than FBP with Hanning window apodization, with T-PLSQ and mFBP exhibiting the best performance.

\bibliographystyle{ieeebib}
\bibliography{/Users/seanrose/PanLab/BibTexLibrary/library,./Extras.bib}

\end{document}